\documentclass[preprint,prd,aps,nofootinbib]{revtex4}
\usepackage{amsmath,graphics}
\begin{document}
\begin{center}
\bibliographystyle{article}
{\Large \textsc{Scalar Casimir effect with non-local boundary conditions}}
\end{center}
\vspace{0.4cm}

\author{Aram Saharian,$^{1,2,3}$ \thanks{
Electronic address: saharian@ictp.it}
Giampiero Esposito$^{4,5}$ \thanks{
Electronic address: giampiero.esposito@na.infn.it}}

\affiliation{${\ }^{1}$Department of Physics, Yerevan State University,
1 Alex Manoogian Street, 375049 Yerevan, Armenia\\
${\ }^{2}$Abdus Salam International Centre for Theoretical Physics,
34014 Trieste, Italy\\
${\ }^{3}$Departamento de F\'{\i}sica-CCEN, Universidade Federal da
Para\'{\i}ba, \\
58.059-970, J. Pessoa, PB C. Postal 5.008, Brazil\\
${\ }^{4}$Istituto Nazionale di Fisica Nucleare, Sezione di Napoli,\\ 
Complesso Universitario di Monte S.
Angelo, Via Cintia, Edificio N', 80126 Naples, Italy\\
${\ }^{5}$Dipartimento di Scienze Fisiche, Complesso Universitario di
Monte S. Angelo,\\ 
Via Cintia, Edificio N', 80126 Naples, Italy}

\begin{abstract}
Non-local boundary conditions have been considered in
theoretical high-energy physics with emphasis on one-loop quantum
cosmology, one-loop conformal anomalies, Bose--Einstein
condensation models and spectral branes.
We have therefore studied the Wightman function, the vacuum expectation
value of the field square and the energy-momentum tensor
for a massive scalar field satisfying non-local
boundary conditions on a single and two parallel plates. Interestingly, 
we find that suitable choices of the kernel in the non-local boundary
conditions lead to forces acting on the plates that can be
repulsive for intermediate distances. 
\end{abstract}
\maketitle
\bigskip
\vspace{2cm}
In our analysis of the Casimir effect for scalar fields \cite{Saha06},
motivated by the work in Refs. \cite{Schr89, Espo99},
we have considered the geometry of two parallel 
plates with non-local boundary conditions 
\begin{equation}
n_{(j)}^{\nu}\partial _{\nu }\varphi (x^{\mu })+\int d\mathbf{x}_{\parallel
}^{\prime }\,f_{j}(|\mathbf{x}_{\parallel }-\mathbf{x}_{\parallel }^{\prime
}|)\varphi (x^{\prime \mu })=0,\;x=a_{j},  
\label{(1)}
\end{equation}
where we use rectangular coordinates $x^{\mu}=(t,x^{1},
\mathbf{x}_{\parallel})$, with $\mathbf{x}_{\parallel}=(x^{2},...,x^{D})$, 
and $n_{(j)}^{\nu}$ is the inward-pointing
unit normal to the boundary at $x=a_{j}$.
For the region between the plates the corresponding
eigenvalues are solutions of the equation \cite{Saha06}
\begin{equation}
\left( z^{2}-c_{1}c_{2}\right) \sin z+\left( c_{1}+c_{2}\right) z\cos z=0,
\label{(2)}
\end{equation}
where the coefficients $c_{j}$ are determined by the 
Fourier transforms $F_{j}$ of the kernel
functions $f_{j}(x_{\parallel})$ in the boundary conditions, i.e.
\begin{equation}
c_{j} \equiv (-1)^{j-1}aF_{j}(k_{\parallel}) \equiv
(-1)^{j-1}a \int d \mathbf{x}_{\parallel}
f_{j}(|\mathbf{x}_{\parallel})
{\rm e}^{{\rm i} \mathbf{k}_{\parallel} \cdot 
\mathbf{x}_{\parallel}}.
\label{(3)}
\end{equation}
The non-local boundary conditions (1) state that the normal derivative at
a given point depends on the values of the field at other points on the
boundary. The properties of the boundary are expressed by the kernel
function $f_{j}$. In a sense, this setting is similar to that in
electrodynamics for the spatial dispersion of the dielectric function 
$\varepsilon$, where $\varepsilon$ depends on the wave vector by virtue
of spatial dispersion. Similarly, our non-local boundary conditions engender
dependence of the coefficient $F_{j}$ in the eigenfunctions on the wave
vector $\mathbf{k}_{\parallel}$.

The evaluation of the corresponding Wightman function is 
based on a variant of the
generalized Abel--Plana summation formula below \cite{Saha06}:
\begin{eqnarray}
\sum_{z=\lambda _{n},iy_{l}}\frac{h(z)}{1+\cos (z+2\alpha _{1})\sin z/z} &=&-
\frac{1}{2}\frac{h(0)}{1-c_{1}^{-1}-c_{2}^{-1}}+\frac{1}{\pi }
\int_{0}^{\infty }dzh(z)  \notag \\
&&+\frac{{\rm i}}{\pi }\int_{0}^{\infty }dt\frac
{h(t{\rm e}^{\pi {\rm i}/2})-h(t{\rm e}^{-\pi {\rm i}/2})}{
\frac{(t-c_{1})(t-c_{2})}{(t+c_{1})(t+c_{2})}{\rm e}^{2t}-1}  \notag \\
&&-\frac{\pi \theta (c_{j})}{2c_{j}}\left[ g_{j}(c_{j}{\rm e}^{\pi
{\rm i}/2})+g_{j}(c_{j}{\rm e}^{-\pi {\rm i}/2})\right],
\label{(4)}
\end{eqnarray}
where $g_{j} \equiv (z^{2}+c_{j}^{2})h(z)$.
The application of this formula has made it possible for us to 
extract from the VEVs the parts
resulting from the single plate and to present the part induced from the
second plate in
terms of integrals exponentially convergent for points away from the
boundary. The Wightman function turns out to be given by \cite{Saha06} 
\begin{eqnarray}
\langle 0|\varphi (x^{\mu })\varphi (x^{\prime \mu })|0\rangle &=&\langle
0_{S}|\varphi (x^{\mu })\varphi (x^{\prime \mu })|0_{S}
\rangle _{j}+\frac{4}{
(2\pi )^{D}}\int d\mathbf{k}_{\parallel }
{\rm e}^{{\rm i}\mathbf{k}_{\parallel} \cdot (\mathbf{
x}_{\parallel}-\mathbf{x}_{\parallel}^{\prime })}  \notag \\
&&\times \int_{a\sqrt{k_{\parallel }^{2}+m^{2}}}^{\infty }dt\frac{\cosh
(tx_{j}+\widetilde{\alpha}_{j})\cosh (tx_{j}^{\prime }
+\widetilde{\alpha}_{j})}{
\frac{(t-c_{1})(t-c_{2})}{(t+c_{1})(t+c_{2})}{\rm e}^{2t}-1}  \notag \\
&&\times \frac{\cosh \left[ (t-t^{\prime })\sqrt{t^{2}/a^{2}-k_{\parallel
}^{2}-m^{2}}\right] }{\sqrt{t^{2}-k_{\parallel }^{2}a^{2}-m^{2}a^{2}}},
\label{(5)}
\end{eqnarray}
having defined ${\widetilde \alpha} \equiv {1\over 2}
\log((t-c_{j})/(t+c_{j}))$.

Moreover, the vacuum stress in the direction orthogonal to the plates
is uniform. This stress determines the vacuum forces acting on the plates,
and the corresponding effective pressure reads as \cite{Saha06}
\begin{eqnarray}
p &=&-\langle 0|T_{1}^{1}|0\rangle =-\frac{2S_{D-1}}{(2\pi )^{D}}
\int_{0}^{\infty }du\,u^{D-2}\int_{\sqrt{u^{2}+m^{2}}}^{\infty }\frac{t^{2}dt
}{\sqrt{t^{2}-u^{2}-m^{2}}}  \notag \\
&&\times \left[ \frac{(t-F_{1}(u))(t+F_{2}(u))}{(t+F_{1}(u))(t-F_{2}(u))}
{\rm e}^{2at}-1\right] ^{-1}.  
\label{(6)}
\end{eqnarray}

We have evaluated numerically the vacuum forces acting on the plates in the
case of the kernel functions \cite{Saha06}
\begin{equation}
f_{j}(x) \equiv f_{0j}e^{-\eta _{j}x}.  
\label{(7)}
\end{equation}
The corresponding Fourier transforms
$F_{j}(k_{\parallel})$ are given by the formulae
\begin{equation}
F_{j}(k_{\parallel})={\eta_{j}F_{1}^{(j)}\over
(1+k_{\parallel}^{2}/\eta_{j}^{2})^{D/2}},
\label{(8)}
\end{equation}
where the parameters $F_{1}^{(j)}$ are defined by
\begin{equation}
F_{1}^{(j)} \equiv 2^{D-1}\pi^{{D\over 2}-1}
\Gamma(D/2){f_{0j}\over \eta^{D}}.
\label{(9)}
\end{equation}
We find that, for the
values $F_{1}^{(1)}\lesssim -1.08$, the vacuum pressure is negative for all
interplate distances and the corresponding vacuum forces are attractive. For
the values $F_{1}^{(1)}>-1.08$ there are two values of the distance between
the plates for which the vacuum forces vanish. These values correspond to
equilibrium positions of the plates. 
Moreover, for values of the distance in the
region between these positions the vacuum forces acting on plates are
repulsive. Thus, the left equilibrium position is unstable and the right
one is locally stable \cite{Saha06}.

It might be interesting to investigate the relation, if any, with the
findings in Ref. \cite{Padm06}, where the authors obtain a repulsive
Casimir force among parallel plates under the assumption of a suitable
ultraviolet cut-off such that the regularized zero-point energy of the
vacuum can be the source of non-vanishing cosmological constant driving
the acceleration of the Universe.
\acknowledgments
The work of A. Saharian has been supported by the INFN, by ANSEF Grant No.
05-PS-hepth-89-70, and in part by the Armenian Ministry of Education and
Science, Grant No. 0124. The work of G. Esposito has been partially
supported by PRIN {\it SINTESI}.

\end{document}